# Effective Yukawa Couplings in Noncompact Lattice QED

M. Göckeler[1,2], R. Horsley[1], P.E.L. Rakow[3] and G. Schierholz[1,4]

[1]Gruppe Theorie der Elementarteilchen,
Höchstleistungsrechenzentrum HLRZ,
c/o Forschungszentrum Jülich, D-52425 Jülich, Germany

[2]Institut für Theoretische Physik, RWTH Aachen,
Physikzentrum, D-52056 Aachen, Germany

[3]Institut für Theoretische Physik, Freie Universität Berlin,
Arnimallee 14, D-14195 Berlin, Germany

[4]Deutsches Elektronen-Synchrotron DESY,
Notkestraße 85, D-22603 Hamburg, Germany

## Abstract

We investigate effective Yukawa couplings of mesons to the elementary fermions in noncompact lattice QED. The couplings are extracted from suitable fermion-antifermion-meson three-point functions calculated by Monte Carlo simulations with dynamical staggered fermions. The scaling behaviour is compatible with expectations from perturbation theory, thus indicating triviality of QED. The lines of constant Yukawa coupling are compared to flows of other quantities. Consistency is seen, at most, for weak coupling.

In strongly coupled lattice QED chiral symmetry is known to be spontaneously broken, whereas an unbroken chiral symmetry is expected in the weak coupling limit [1]. This finding immediately raises questions like: Can a nontrivial continuum limit be defined at the corresponding critical point? If so, what is the nature of the associated continuum theory? Several groups have investigated these problems by means of extensive Monte Carlo simulations of lattice QED using various techniques (see, e.g., [2-9]). Yet the physical interpretation of the data is still controversial (see also [10]).

Correlation functions of composite fermion-antifermion operators show evidence for at least two fermion-antifermion (bound) states ("meson" states): a pseudoscalar state denoted by P (the Goldstone boson associated with the spontaneous symmetry breaking) and a scalar state (S) [3, 11, 6]. In this letter we study fermion-antifermion-meson three-point functions in order to determine (effective) Yukawa couplings of P and S to the fermions. The investigation should provide additional information about the lattice model. In particular, if it has a trivial continuum limit, the couplings should tend to zero as one approaches the critical point.

We have performed simulations with dynamical staggered fermions using the noncompact formulation of the gauge field action. So the total action is given by $S = S_G + S_F$ with

$$S_G = \tfrac{1}{2}\beta \sum_{x,\mu<\nu} F_{\mu\nu}^2(x), \tag{1}$$

$$S_F = \sum_x \left\{ \tfrac{1}{2} \sum_\mu \eta_\mu(x) \left( \overline{\chi}(x) e^{iA_\mu(x)} \chi(x+\hat{\mu}) - \overline{\chi}(x+\hat{\mu}) e^{-iA_\mu(x)} \chi(x) \right) + m\overline{\chi}(x)\chi(x) \right\} \tag{2}$$

where $\beta = 1/e^2$ ($e$ = bare charge) and

$$\eta_\mu(x) = (-1)^{x_1 + \cdots + x_{\mu-1}}, \tag{3}$$

$$F_{\mu\nu}(x) = A_\mu(x) + A_\nu(x+\hat{\mu}) - A_\mu(x+\hat{\nu}) - A_\nu(x). \tag{4}$$

The data in this paper were obtained on an $L_s^3 \times L_t$ lattice with $L_s = L_t = 12$. The boundary conditions are periodic for the gauge field $A_\mu$. For the fermion field $\chi, \overline{\chi}$ we have chosen periodic (antiperiodic) spatial (temporal) boundary conditions. For more details on the simulation see ref. [6].

In order to calculate fermionic observables like the fermion propagator or the fermion-antifermion-meson three-point functions, which are the subject of this paper, we have to fix the gauge. We choose to work with the Landau gauge,

$$\sum_\mu (A_\mu(x) - A_\mu(x-\hat{\mu})) = 0, \tag{5}$$

which has the advantage that it can be implemented exactly in the noncompact formulation. Unfortunately, the zero-momentum mode of the gauge field does not average to zero in our ensembles. To cope with this problem we divide our data into blocks of, say, 20 successive configurations, on which this mode is approximately constant, and perform our analysis in each block separately taking the zero-momentum mode into account explicitly (cf. [6]). The averages and errors are then obtained from these block results.

As we have argued in ref. [6], the renormalized charge vanishes as one approaches the critical point $\beta_c \approx 0.19$, $m = 0$. Moreover, it turns out that the chiral phase transition is reasonably well described by logarithmically improved mean field theory, i.e., the equation of state can be derived from an effective linear $\sigma$-model [6, 12]. Within such a model, the effective quartic coupling $\lambda$ is driven to zero in the continuum limit according to a (one-loop) renormalization group equation of the form

$$\frac{\mathrm{d}\lambda}{\mathrm{d}t} = c\lambda^2 \,, t = \ln \sigma \,, \tag{6}$$

where $\sigma = \langle \overline{\chi}\chi \rangle$ is the chiral condensate and $c > 0$. From the solution of this equation one sees that $\lambda \sim |\ln \sigma|^{-1}$ as $\sigma \to 0$. On the other hand, for a Yukawa coupling $g$, which we are investigating here, one finds in leading order [13]

$$\frac{\mathrm{d}g}{\mathrm{d}t} = c'g^3 \,, \quad c' > 0 \,, \tag{7}$$

so that

$$\frac{1}{g^2} = -2c't + \mathrm{const.} \tag{8}$$

Hence the triviality scenario leads to the expectation that $g$ vanishes like

$$g \sim |\ln \sigma|^{-1/2} \,. \tag{9}$$

We shall extract our effective Yukawa couplings from three-point functions of the form

$$G_3^{(\alpha)}(\vec{p}, \vec{q}; t_0, t_1) = \sum_{\vec{x}, \vec{x}'} \mathrm{e}^{2\mathrm{i}\vec{q}\cdot\vec{x}' - 2\mathrm{i}\vec{p}\cdot\vec{x}} \langle \overline{\chi}(2\vec{x}, t_1) M_\alpha(\vec{p} - \vec{q}, 0)\chi(2\vec{x}', t_0)\rangle_c \,. \tag{10}$$

Here $\langle \cdots \rangle_c$ denotes the fermion-line connected part of the expectation value, the lattice vectors $\vec{x}, \vec{x}'$ label spatial cubes of size $2^3$, and $M_\alpha(\vec{p}_0, t)$ is the standard pseudoscalar ($\alpha = P$) or scalar ($\alpha = S$) meson operator:

$$M_\alpha(\vec{p}_0, t) = \sum_{\vec{y}} \mathrm{e}^{\mathrm{i}\vec{p}_0\cdot\vec{y}} \varphi_\alpha(\vec{y})\overline{\chi}(\vec{y}, t)\chi(\vec{y}, t) \tag{11}$$

with $\varphi_P(\vec{y}) = (-1)^{y_1+y_2+y_3}, \varphi_S(\vec{y}) = 1$. We have calculated the three-point function (10) for the following values of the spatial fermion momenta $\vec{p}, \vec{q}$:

$$(\vec{p}, \vec{q}) = (\vec{0}, \vec{0}), (\frac{2\pi}{L_s}\vec{e}_3, \vec{0}), (\frac{2\pi}{L_s}\vec{e}_3, \frac{2\pi}{L_s}\vec{e}_3). \tag{12}$$

The spatial meson momentum is then given by $\vec{p}_0 = \vec{p} - \vec{q}$.

As is well known, the operators (11) couple not only to the state one is interested in but also to its "parity partner" (and higher excitations). Having computed the three-point function (10) for arbitrary times $t_0, t_1$ we can, however, project onto a single parity by Fourier transforming the $t_0, t_1$ dependences:

$$\sum_{t_1, t_0} e^{-ip_4 t_1 + iq_4 t_0} G_3^{(\alpha)}(\vec{p}, \vec{q}; t_0, t_1)$$
$$= \frac{1}{L_t} \sum_{t_1, t_0} e^{-ip_4 t_1 + iq_4 t_0} \sum_{\vec{x}, \vec{x}'} e^{2i\vec{q}\cdot\vec{x}' - 2i\vec{p}\cdot\vec{x}} \langle \overline{\chi}(2\vec{x}, t_1) \sum_t e^{i(p_4-q_4)t} M_\alpha(\vec{p}-\vec{q}, t) \chi(2\vec{x}', t_0) \rangle_c. \tag{13}$$

Since the Goldstone boson is represented by the component of $M_P$ that "oscillates" in time (see, e.g., [14]), we must choose $p_4 - q_4 = \pi$ in order to study the Yukawa coupling of the pseudoscalar. In the case of the scalar, on the other hand, we have to take $p_4 - q_4 = 0$.

We define effective Yukawa couplings $g_\alpha$ by comparing the Monte Carlo data for the three-point functions with tree-level formulas derived from an effective lattice action which describes the interaction between staggered fermions $\chi, \overline{\chi}$ and a (pseudo-)scalar field $\Phi$ via coupling terms of the form

$$-g_S \sum_x \Phi(x)\overline{\chi}(x)\chi(x) \quad \text{(scalar)}, \tag{14}$$

$$-g_P \sum_x \Phi(x)\overline{\chi}(x)\chi(x)(-1)^{x_1+\cdots+x_4} \quad \text{(pseudoscalar)}. \tag{15}$$

In the tree-level formulas for the three-point functions we replace the free propagators by the full propagators calculated in the simulations. Furthermore, we include the appropriate wave function renormalizations $Z_F$ (for the fermion field) and $Z_\alpha$ (for the mesons). In this way we arrive at the following equation defining the momentum-dependent Yukawa couplings $g_\alpha(p,q)$:

$$\sum_{t_1,t_0} e^{-ip_4 t_1 + iq_4 t_0} G_3^{(\alpha)}(\vec{p},\vec{q};t_0,t_1)$$
$$= g_\alpha(p,q) Z_F(\vec{p})^{-\frac{1}{2}} Z_F(\vec{q})^{-\frac{1}{2}} Z_\alpha(\vec{p}-\vec{q})^{-\frac{1}{2}}$$
$$\times \frac{8}{L_s^3} \theta_\alpha \sum_t e^{i(p_4-q_4)t} \frac{1}{L_s^3} \langle M_\alpha(\vec{q}-\vec{p},t) M_\alpha(\vec{p}-\vec{q},0)\rangle_c \qquad (16)$$
$$\times \sum_{\vec{\omega}} \bar{\theta}_\alpha(\vec{\omega}) e^{-i(\vec{p}-\vec{q})\cdot\vec{\omega}} \left\{ \sum_{t_0} \eta_4(\vec{\omega})^{t_0} e^{iq_4 t_0} \sum_{\vec{x},\vec{x}'} e^{2i\vec{q}\cdot(\vec{x}-\vec{x}')} \langle \chi(2\vec{x}+\vec{\omega},t_0) \overline{\chi}(2\vec{x}',0)\rangle \right\}$$
$$\times \left\{ \sum_{t_1} \eta_4(\vec{\omega})^{t_1} e^{i(p_4+\pi)t_1} \sum_{\vec{x},\vec{x}'} e^{2i\vec{p}\cdot(\vec{x}-\vec{x}')} \langle \chi(2\vec{x}+\vec{\omega},t_1) \overline{\chi}(2\vec{x}',0)\rangle \right\}^*,$$

where
$$\theta_S = -1, \theta_P = 1, \qquad (17)$$
$$\bar{\theta}_S(\vec{\omega}) = (-1)^{\omega_1+\omega_2+\omega_3}, \bar{\theta}_P(\vec{\omega}) = 1,$$

and $\vec{\omega}$ runs over the 8 three-vectors with components 0 or 1, i.e. $\vec{\omega}$ labels the corners of a spatial unit cube.

The wave function renormalizations $Z_F, Z_\alpha$ are obtained in the usual way by comparing the one-particle contributions in the full propagators with the corresponding free propagators. For example, writing the one-particle contribution to the meson propagator

$$\frac{1}{L_s^3} \langle M_\alpha(-\vec{p},t) M_\alpha(\vec{p},0)\rangle_c \qquad (18)$$

in the form

$$A_\alpha(\vec{p}) \left( e^{-E_\alpha(\vec{p})t} + e^{-E_\alpha(\vec{p})(L_t-t)} \right) \qquad (19)$$

we have

$$Z_\alpha(\vec{p}) = 2|A_\alpha(\vec{p})| \left(1 - e^{-L_t E_\alpha(\vec{p})}\right) \sinh E_\alpha(\vec{p}) \qquad (20)$$

(see, e.g., ref.[15]). In the case of the fermion wave function renormalization we compare with the propagator in a constant background potential [6].

Let us now discuss the results. One first observes that $g_\alpha(p,q)$ is essentially real: The imaginary part is relatively small and noisy, for special values of the momenta it is even exactly zero. Hence we shall disregard it and consider $g_\alpha$ as a real quantity. For obtaining a final unique answer it would be desirable to send all momenta to zero. However, the antiperiodic temporal boundary conditions for the fermions would allow us only to give the mesons momentum zero by choosing $\vec{p}=\vec{q}, p_4 = q_4$ for the scalar and $p_4 = q_4 + \pi$ (mod $2\pi$) for the pseudoscalar. Admitting for $(\vec{p}, \vec{q})$ the values (12) we give $|q_4|$ the smallest possible value and average over $q_4 = \pm \pi/L_t$:

$$\bar{g}_S(\vec{p}, \vec{q}) = \tfrac{1}{2} \sum_{q_4 = \pm \pi/L_t} g_S((\vec{p}, q_4), (\vec{q}, q_4)), \tag{21}$$

$$\bar{g}_P(\vec{p}, \vec{q}) = \tfrac{1}{2} \sum_{q_4 = \pm \pi/L_t} g_P((\vec{p}, q_4 + \pi), (\vec{q}, q_4)). \tag{22}$$

For the error we take the largest error of a single value.

In fig. 1 we plot $\bar{g}_P(\vec{p}, \vec{q})^{-2}$ versus $|\ln \sigma|$ as suggested by eq. (8). (Note that in the preliminary presentation of our results in [16] we have erroneously shown the average of $g_P((\vec{0}, q_4), (\vec{0}, q_4))$.) Although these numbers were obtained at different bare couplings $\beta$ ($0.17 \leq \beta \leq 0.22$) and different bare masses $m$ ($0.02 \leq m \leq 0.09$), they fall into a narrow band, which with some optimism could even be considered as a single straight line in accordance with the triviality scenario (cf. eq. (8)). We consider these data as a qualitative confirmation of the triviality hypothesis. The results for the coupling of the scalar shown in fig. 2 point into the same direction.

However, several caveats have to be kept in mind. We have neglected the fermion-line disconnected contributions to the three-point functions and to the meson propagators, since they are difficult to calculate. These contributions could be especially important in the case of the scalar, which is a flavour singlet. Furthermore it is not clear, how our Yukawa couplings are affected by the fact that for some of our data points the mesons are heavier than two fermions and hence might correspond to resonances rather than stable particles in the continuum limit. Another problem concerns the definition of the wave function renormalizations. Since they are extracted from fits of free propagators to the measured two-point functions, they correspond to on-shell renormalization, whereas the effective Yukawa couplings are defined at fixed momenta ("fixed" in lattice units, the physical scale varies from simulation point to simulation point). The weak momentum dependence of the results may however indicate that this problem is not too severe.

With the effective Yukawa couplings at hand we can draw new flow diagrams, i.e. we can construct the lines of constant $\bar{g}_S$, $\bar{g}_P$ in the $\beta$-$m$ plane. These are to be compared with the lines of constant renormalized charge $e_R$ or constant mass ratios [4, 6]. In fig. 3 we show the curves of constant $\bar{g}_P(\vec{0}, \vec{0})$ together with the lines of constant $e_R$. We recall that the curves of constant $e_R > 0$ end on the line of first-order chiral phase transitions $m = 0, \beta < \beta_c$, while only the curve $e_R = 0$ flows into the critical point [6], in accordance with triviality. The lines of constant $\bar{g}_P(\vec{0}, \vec{0})$ look rather similar, although truly parallel flows seem to be possible only in the lower right hand corner, i.e. for small values of $e_R$ and $\bar{g}_P$. This suggests that also in the case of the Yukawa couplings only the curve of vanishing coupling flows into the critical point as perturbation theory would predict.

In fig. 4 we compare with the lines of constant $m_R/m_P$ where $m_R$ is the renormalized fermion mass and $m_P$ the pseudoscalar mass. Here we observe a completely different flow pattern in the left part of the diagram, whereas there is a tendency towards parallel flows in the region of small $e_R$. The corresponding plots for $\bar{g}_S$ look qualitatively similar.

In conclusion, these results are in agreement with our previous findings [4, 6] indicating that lines of (approximately) constant physics can only be expected for small couplings. To achieve renormalizability for larger couplings it would seem to be necessary to include further relevant operators in the action, such as four-fermion terms.

# ACKNOWLEDGEMENTS

This work was supported in part by the Deutsche Forschungsgemeinschaft. The numerical computations were performed on the Cray Y-MP in Jülich with time granted by the Scientific Council of the HLRZ. We wish to thank both institutions for their support.

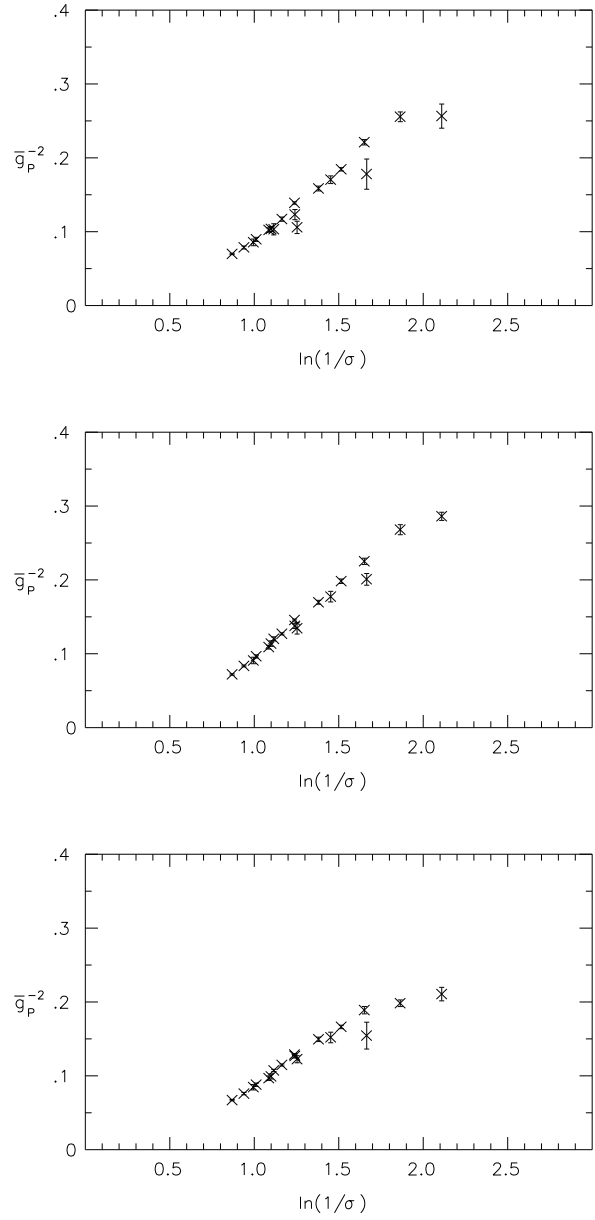

Figure 1: $\bar{g}_P(\vec{p},\vec{q})^{-2}$ versus $\ln(1/\sigma)$ for $(\vec{p},\vec{q}) = (\vec{0},\vec{0}),(\frac{2\pi}{L_s}\vec{e}_3,\frac{2\pi}{L_s}\vec{e}_3),(\frac{2\pi}{L_s}\vec{e}_3,\vec{0})$ (top to bottom).

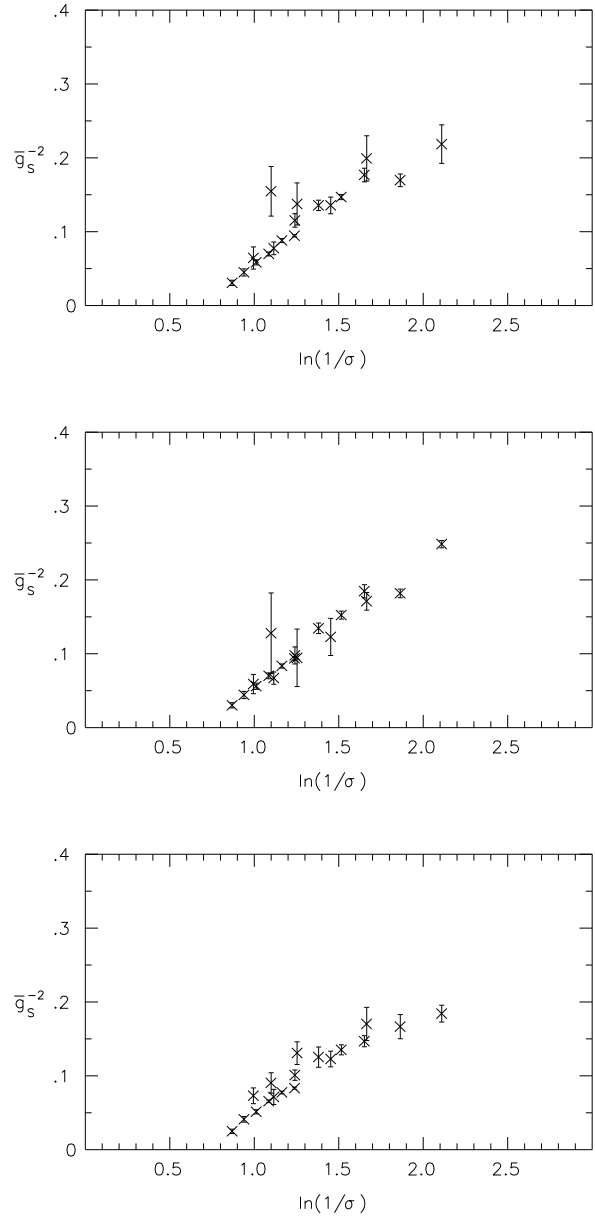

Figure 2: $\bar{g}_S(\vec{p},\vec{q})^{-2}$ versus $\ln(1/\sigma)$ for $(\vec{p},\vec{q}) = (\vec{0},\vec{0}), (\frac{2\pi}{L_s}\vec{e}_3, \frac{2\pi}{L_s}\vec{e}_3), (\frac{2\pi}{L_s}\vec{e}_3, \vec{0})$ (top to bottom).

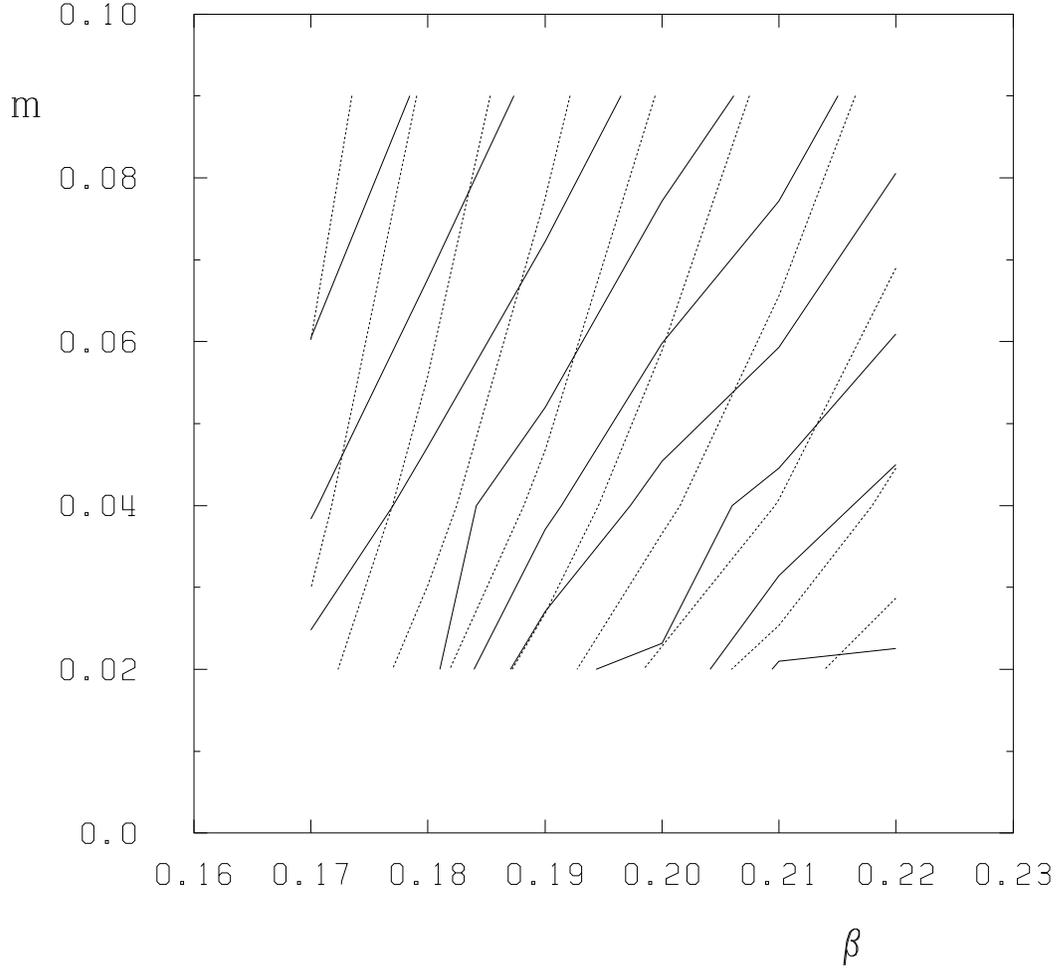

Figure 3: Curves of constant $\bar{g}_P(\vec{0}, \vec{0})$ (full lines) and curves of constant renormalized charge (dashed lines) with $\bar{g}_P(\vec{0}, \vec{0})$ ranging from 2.0 (lower right-hand corner) to 3.6 (upper left-hand corner) in steps of 0.2 and $e_R^2$ ranging from 2.8 (lower right-hand corner) to 4.6 (upper left-hand corner) in steps of 0.2.

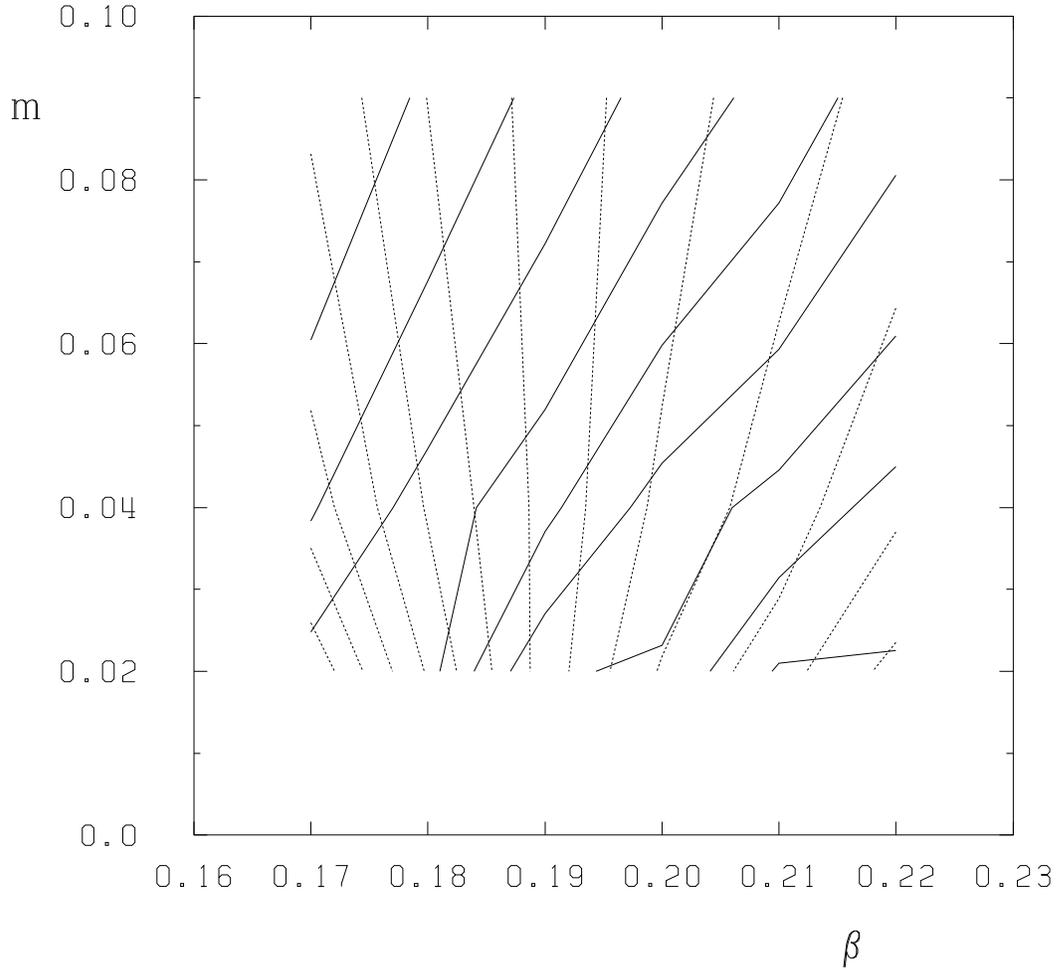

Figure 4: Curves of constant $\bar{g}_P(\vec{0},\vec{0})$ (full lines) and curves of constant $m_R/m_P$ (dashed lines) with $\bar{g}_P(\vec{0},\vec{0})$ ranging from 2.0 (lower right-hand corner) to 3.6 (upper left-hand corner) in steps of 0.2 and $m_R/m_P$ ranging from 0.4 (lower right-hand corner) to 1.6 (lower left-hand corner) in steps of 0.1.